\documentclass[3p,times,procedia]{elsarticle}
\usepackage{nupha_ecrc}

\volume{00}

\firstpage{1}

\journalname{Nuclear Physics A}

\runauth{}

\jid{nupha}

\jnltitlelogo{Nuclear Physics A}

\usepackage{amssymb}
\usepackage[figuresright]{rotating}

\begin{document}

\begin{frontmatter}

%% Title, authors and addresses

%% use the tnoteref command within \title for footnotes;
%% use the tnotetext command for the associated footnote;
%% use the fnref command within \author or \address for footnotes;
%% use the fntext command for the associated footnote;
%% use the corref command within \author for corresponding author footnotes;
%% use the cortext command for the associated footnote;
%% use the ead command for the email address,
%% and the form \ead[url] for the home page:
%%
%% \title{Title\tnoteref{label1}}
%% \tnotetext[label1]{}
%% \author{Name\corref{cor1}\fnref{label2}}
%% \ead{email address}
%% \ead[url]{home page}
%% \fntext[label2]{}
%% \cortext[cor1]{}
%% \address{Address\fnref{label3}}
%% \fntext[label3]{}

%% Instructions from Editor: Please use the following \dochead only in the preprint version (e-print arXiv etc.);
%% use empty \dochead{} when submitting to Nuclear Physics A!
\dochead{XXVIth International Conference on Ultrarelativistic Nucleus-Nucleus Collisions\\ (Quark Matter 2017)}
%\dochead{}
%% Use \dochead if there is an article header, e.g. \dochead{Short communication}
%% \dochead can also be used to include a conference title, if directed by the editors
%% e.g. \dochead{17th International Conference on Dynamical Processes in Excited States of Solids}

\title{Beam energy dependence of $d$ and $\bar{d}$ productions in Au+Au collisions at RHIC}

%% use optional labels to link authors explicitly to addresses:
%% \author[label1,label2]{<author name>}
%% \address[label1]{<address>}
%% \address[label2]{<address>}

\author{Ning Yu (for the STAR Collaboration)}

\address{College of Physical Science and Technology, Central China Normal University, Wuhan, China}

\begin{abstract}

Light nuclei have much smaller binding energy compared to the temperature of the system created in heavy-ion collisions. Consequently, the distributions of light nuclei can be used to probe the freeze-out properties, such as correlation volume and local baryon density of the medium created in high-energy nuclear collisions. In this paper, we report the results of deuteron and anti-deuteron production in Au+Au collision at $\sqrt{s_\mathrm{NN}}$ = 7.7, 11.5, 14.5, 19.6, 27, 39, 62.4, and 200 GeV, measured by STAR at RHIC. The collision energy, centrality and transverse momentum dependence of the coalescence parameter $B_2$ for deuteron and anti-deuteron production are discussed. We find the values of $B_2$ for anti-deuteron are systematically lower than those for deuterons indicating the correlation volume of anti-baryon are larger than that of baryon. In addition, the values of $B_2$ are found to decrease with increasing collision energy and reach a minimum around $\sqrt{s_\mathrm{NN}}$ = 20 GeV implying a change of the equation of state of the medium in these collisions.
\end{abstract}

\begin{keyword}
%% keywords here, in the form: keyword \sep keyword
Deuteron \sep Coalescence parameters
%% MSC codes here, in the form: \MSC code \sep code
%% or \MSC[2008] code \sep code (2000 is the default)

\end{keyword}

\end{frontmatter}

%%
%% Start line numbering here if you want
%%
% \linenumbers

%% main text
\section{Introduction}
\label{}
The formation of nuclei and anti-nuclei in relativistic heavy-ion collisions is through coalescence of produced or participant nucleons and produced anti-nucleons~\cite{PhysRevLett.37.667,PhysRevC.59.1585,SATO1981153}. Since the binding energy of light nuclei is small ($\sim$2.2 MeV for $d$ ($\bar{d}$) and $\sim$7.7 MeV for $^3$He), they can not survive when the temperature is high. They might break apart and be created again by final-state coalescence when the temperature is low.

In such a coalescence picture~\cite{PhysRev.129.836}, the invariant yields of light nuclei with charge $Z$ and atomic mass number $A$ can be described as the product of the yields of their constituents (protons and neutrons) and through a coefficient $B_A$, the so-called coalescence parameter,
\begin{equation}
	{E_A}\frac{{{d^3}{N_A}}}{{{d^3}{p_A}}} = {B_A}{\left( {{E_p}\frac{{{d^3}{N_p}}}{{{d^3}{p_p}}}} \right)^Z}{\left( {{E_n}\frac{{{d^3}{N_n}}}{{{d^3}{p_n}}}} \right)^{A - Z}} \approx {B_A}{\left( {{E_p}\frac{{{d^3}{N_p}}}{{{d^3}{p_p}}}} \right)^A}
\end{equation}
where  $p_A = Ap_p$. This is applicable assuming that the distributions of neutrons and protons are the same. The coalescence parameter $B_A$ reflects the probability of nucleon coalescence, which is related to the local nucleon density.

The production of light nuclei can also be described by thermodynamic models~\cite{Andronic2011203,PhysRevC.84.054916,0954-3899-38-12-124081}, in which chemical equilibrium among protons, neutrons and light nuclei is achieved. The effective volume of the nuclear matter at the time of condensation of nucleons into nuclear clusters, also called ``nucleon correlation volume $V_{\mathrm{eff}}$'', is related to the coalescence parameter $B_A$,
\begin{equation}
    {B_A}\propto V_{\mathrm{eff}}^{1-A}
\end{equation}

For deuteron production~\cite{PhysRevC.60.031901,E802}, ${B_2}\propto 1/V$. The production of light nuclei can provide us with useful information on nucleon density and correlation of nucleons.

%%\section{Analysis method}
%%The analysis is based on data taken at STAR in Au+Au collisions at $\sqrt{s_\mathrm{NN}} =$ 7.7, 11.5, 14.5, 19.6, 27, 39, 62.4, and 200 GeV from year 2010, 2011, and 2014. In the analysis, we used the Time Projection Chamber (TPC) and Time of Flight (TOF) detectors to identify the light nuclei. In TPC, particle identification is accomplished by measuring the ionization energy loss ($dE/dx$). A new variable $z$ is useful to separate out the nuclei from other charged tracks, which is defined as
%%\begin{equation}\label{zfun}
%%z=\mathrm{log}\frac{\langle dE/dx\rangle_\mathrm{measure}}{\langle dE/dx\rangle_\mathrm{Bichsel}}
%%\end{equation}
%%where $\langle dE/dx\rangle_\mathrm{Bichsel}$ is the predicted $dE/dx$ for a fixed particle type of Bichsel model~\cite{Bichsel2006154}. TOF can enhance particle identification for the tracks in higher $p_T (>1.0$GeV/$c)$ by $m^2$.
%%\begin{equation}\label{m2}
%%m^2=p^2\left(\frac{c^2\tau^2}{L^2}-1\right)
%%\end{equation}
%%where, $p$, $t$, $L$, and $c$ are the momentum, time-of-travel by particle, path length, and velocity of light, respectively. The TPC detector acceptance and tracking efficiency, energy loss, and absorption corrections are obtained from Monte Carlo embedding method. The TOF matching efficiency are obtained from the ratios of TOF matched tracks to TPC tracks with the same cuts.

\section{Results and discussions}
\begin{figure}[!h]
\centering\scalebox{0.372}[0.372]{\includegraphics{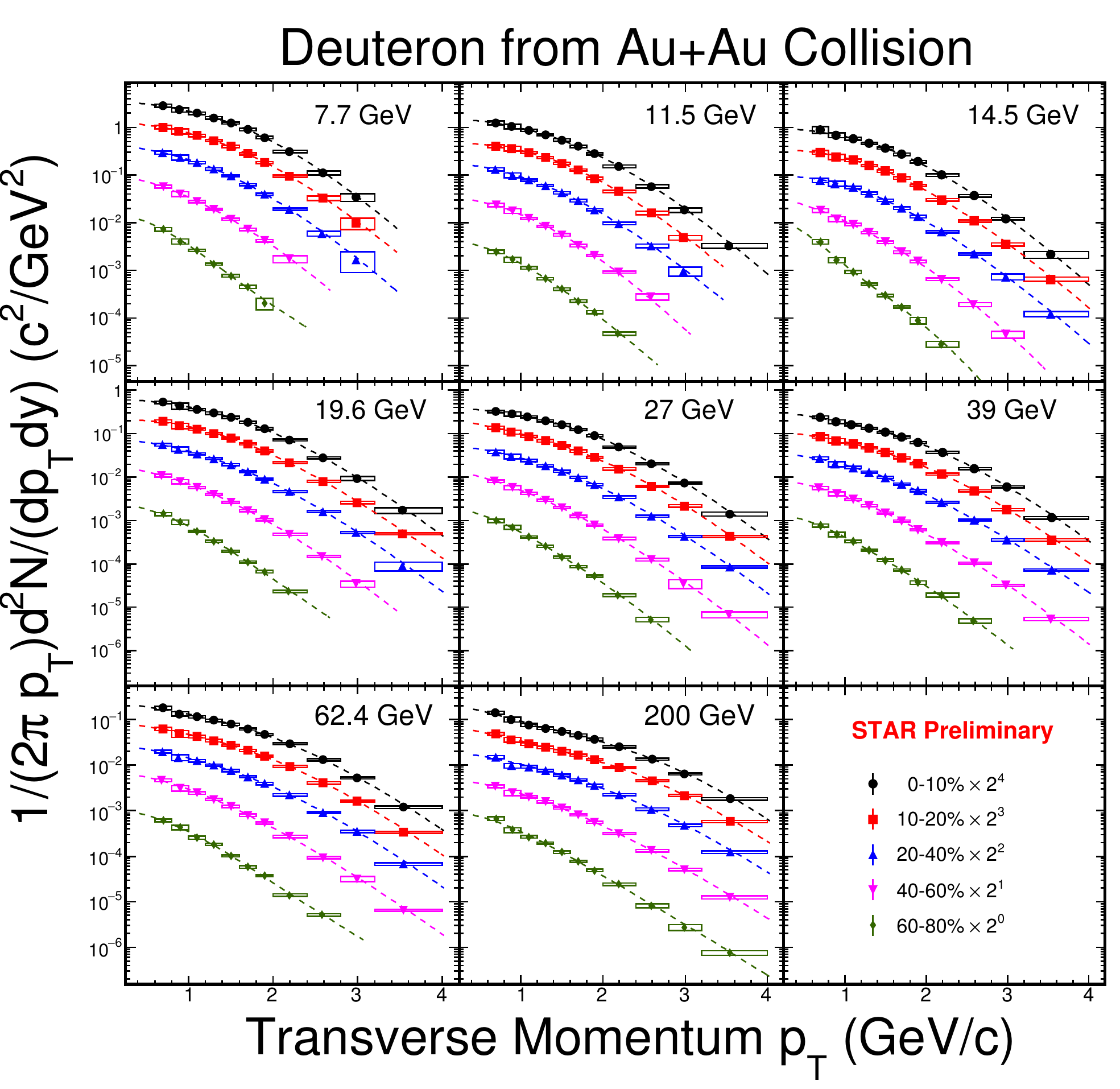}}
\scalebox{0.372}[0.372]{\includegraphics{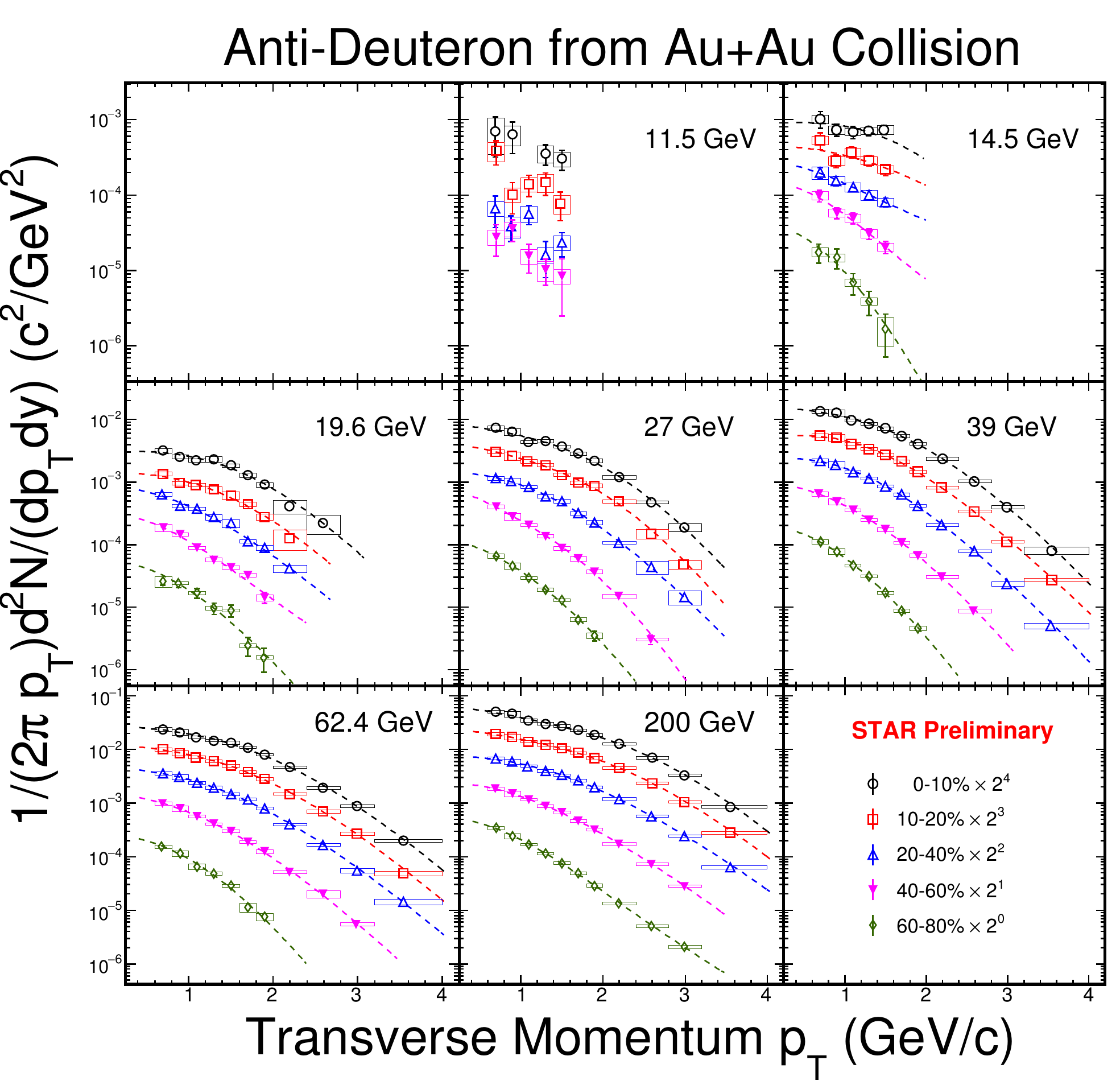}}
\caption{\label{dspectra} Mid-rapidity ($|y| <$ 0.3) transverse momentum spectra for $d$ in Au+Au collisions at $\sqrt{s_\mathrm{NN}} =$ 7.7 $\sim$ 200 GeV for 0-10\%, 10-20\%, 20-40\%, 40-60\%, and 60-80\% centralities. The dash lines are the individual fits with blast-wave functions. The boxes show the systematic error and vertical lines show the statistical error separately.}
\end{figure}

The results presented here are based on RHIC Beam Energy Scan (BES) and are obtained for Au+Au collisions at $\sqrt{s_\mathrm{NN}} =$ 7.7$\sim$200 GeV at mid-rapidity ($|y|<$0.3) using both STAR Time Projection Chamber (TPC) and Time Of Flight (TOF) detectors. The transverse momentum spectra of identified $d$ (left panel) and $\bar{d}$ (right panel) at five collision centrality intervals in Au+Au collisions are shown in Fig.~\ref{dspectra}. Due to the limited statistics and low $dN/dy$, there are not enough candidates to show the $\bar{d}$ spectra of 7.7 GeV. The spectra show a hardening with increasing multiplicity/centrality and are fitted with individual blast-wave functions. The results of the blast-wave fitting are shown as dashed lines in the Fig.~\ref{dspectra}.

\begin{figure}[!ht]
\scalebox{0.375}[0.375]{\includegraphics{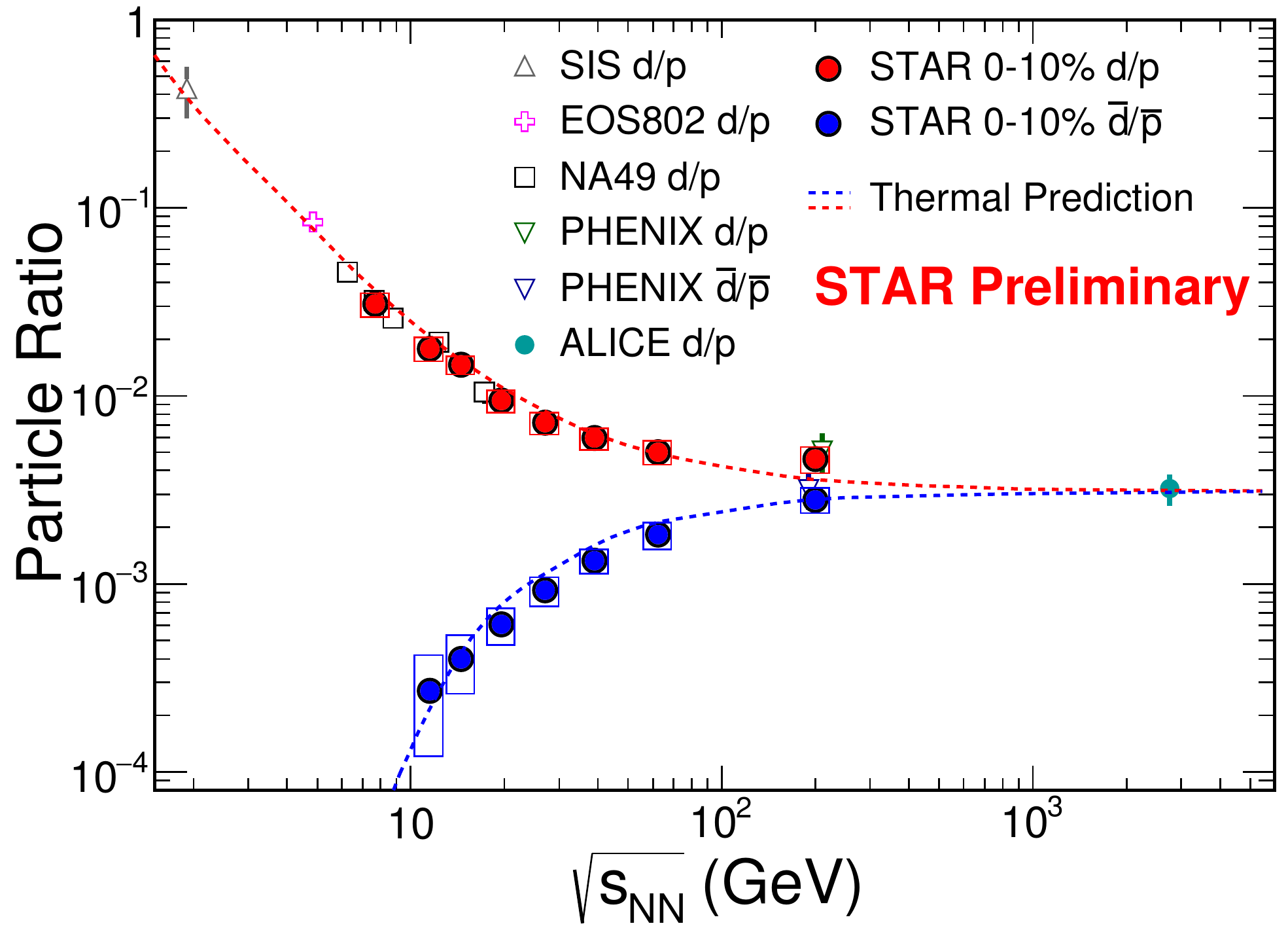}}
\scalebox{0.375}[0.375]{\includegraphics{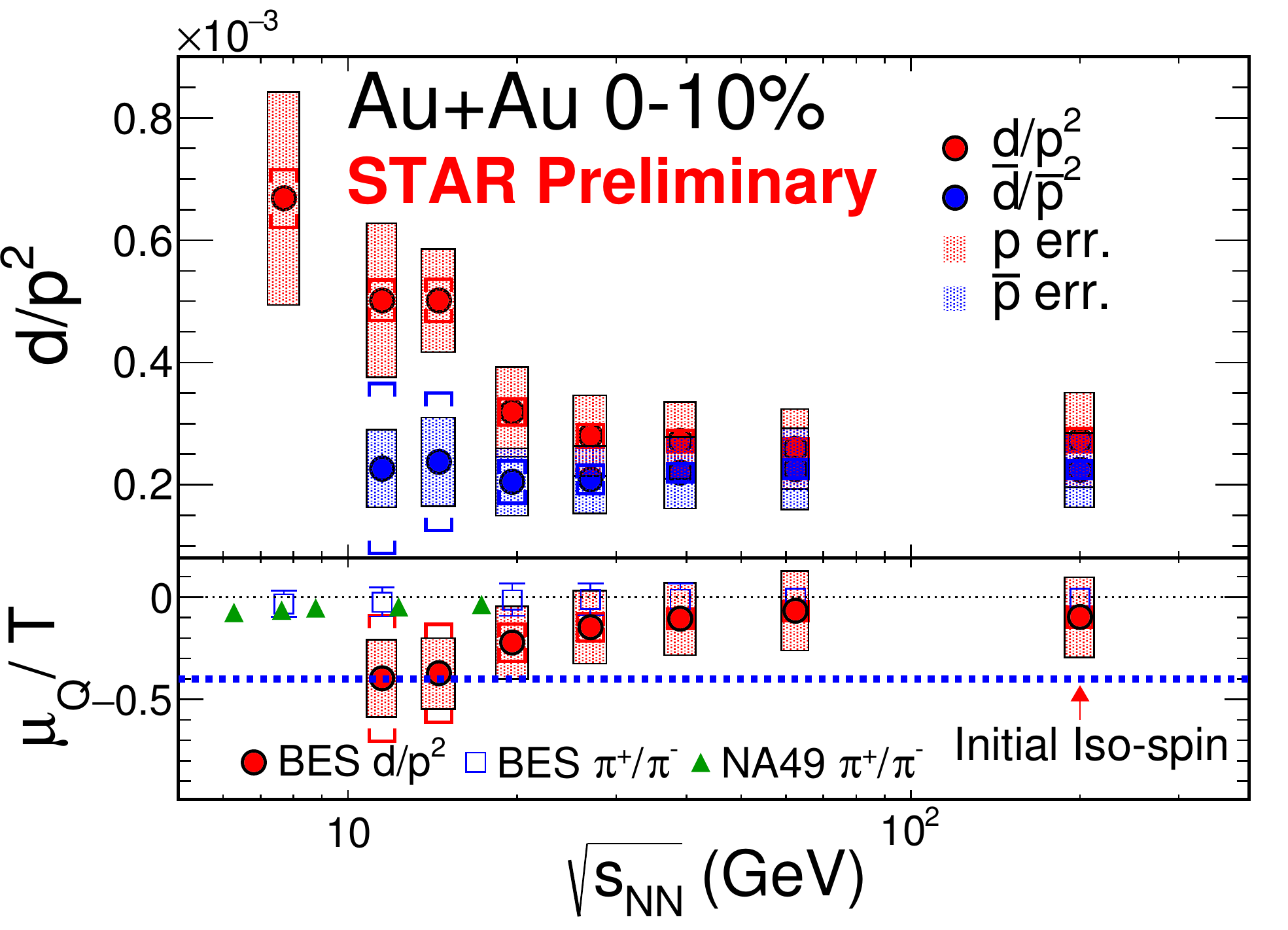}}
\caption{\label{pratios} Left panel : $d/p$ (red circles) and $\bar{d}/\bar{p}$ (blue circles) ratios for Au+Au 0-10\% central collision at BES energies. The dotted lines are thermal model predictions. Right top panel : $d/p^2$ (red circles) and $\bar{d}/\bar{p}^2$ (blue circles) ratios for Au+Au 0-10\% central collision at BES energies. Right bottom panel : Extracted $\mu_Q/T$ from BES $d/p^2$ (red circles), BES $\pi^{+}/\pi^{-}$, and NA49 $\pi^{+}/\pi^{-}$~\cite{NA49}. The blue dotted-line represents the initial iso-spin for Au+Au collision. The brackets show systematic error and vertical lines show the statistical error separately. The boxes are errors for proton and anti-proton.}
\end{figure}

The $p_T$-integrated $dN/dy$ values are obtained by sum of the experimental measured data points and the blast-wave results for the unmeasured $p_T$ regions. Left panel of Fig.~\ref{pratios} shows the results of $d/p$ and $\bar{d}/\bar{p}$ $dN/dy$ ratios in 0-10\% Au+Au collision at BES energies and compared with the central collision results of SIS, AGS, SPS, RHIC PHENIX, and LHC~\cite{E802,PhysRevC.59.1663,NA49,PhysRevLett.94.122302,ALICE}. The proton and anti-proton are corrected for the weak decay feed-down. The dash lines are results from thermal model calculation employing parameters established from the analysis of light hadron production in relativistic nuclear collisions~\cite{Andronic2011203}. The thermal results fit to the ratios of $d/p$ and $\bar{d}/\bar{p}$ well.

To study the isospin effect in heavy-ion collision, the $d/p^2$ and $\bar{d}/\bar{p}^2$ ratios in 0-10\% Au+Au collision are shown in right panel of Fig.~\ref{pratios}. In thermal model with grand canonical ensemble, we can find that

\begin{equation}
\frac{\bar{d}/\bar{p}^2}{d/p^2}=e^{\frac{2\mu_Q}{T}}
\end{equation}
If the isospin effect can be ignored, these $\mu_Q$ should be zero. However, in the right bottom panel of Fig.~\ref{pratios}, it is shown that $\mu_Q/T\neq0$ in central Au+Au collision below 20 GeV. $\mu_Q/T=-0.42$ from initial $p/n$ ratio is shown as a blue dotted line. The $\mu_Q/T$ seems to increase with collision energy and reach saturation ($=0$) at high energy, because almost all the particles and anti-particles are produced in the mid-rapidity at high energies. The $\mu_Q/T$ can also be obtained from the $\pi^{+}/\pi^{-}$ ratio as
\begin{equation}
\frac{\mu_Q}{T}=\frac{1}{2}\ln\left(\frac{\pi^{+}}{\pi^{-}}\right)
\end{equation}

In the right bottom panel of Fig.~\ref{pratios}, the $\mu_Q/T$ from RHIC-STAR BES and SPS NA49 $\pi^{+}/\pi^{-}$ results are shown. It is found that, at $\sqrt{s_\mathrm{NN}} <$ 20 GeV, the $-\mu_Q/T$ from $\pi^{+}/\pi^{-}$ are larger than those from $(\bar{d}/\bar{p}^2)/(d/p^2)$, which may indicate that there are some residual fragments which do not fragment to nucleons or quarks at low collision energies. At these low collision energies, the baryon degree of freedom is dominant.

\begin{figure}[!ht]
\centering\scalebox{0.6}[0.6]{\includegraphics{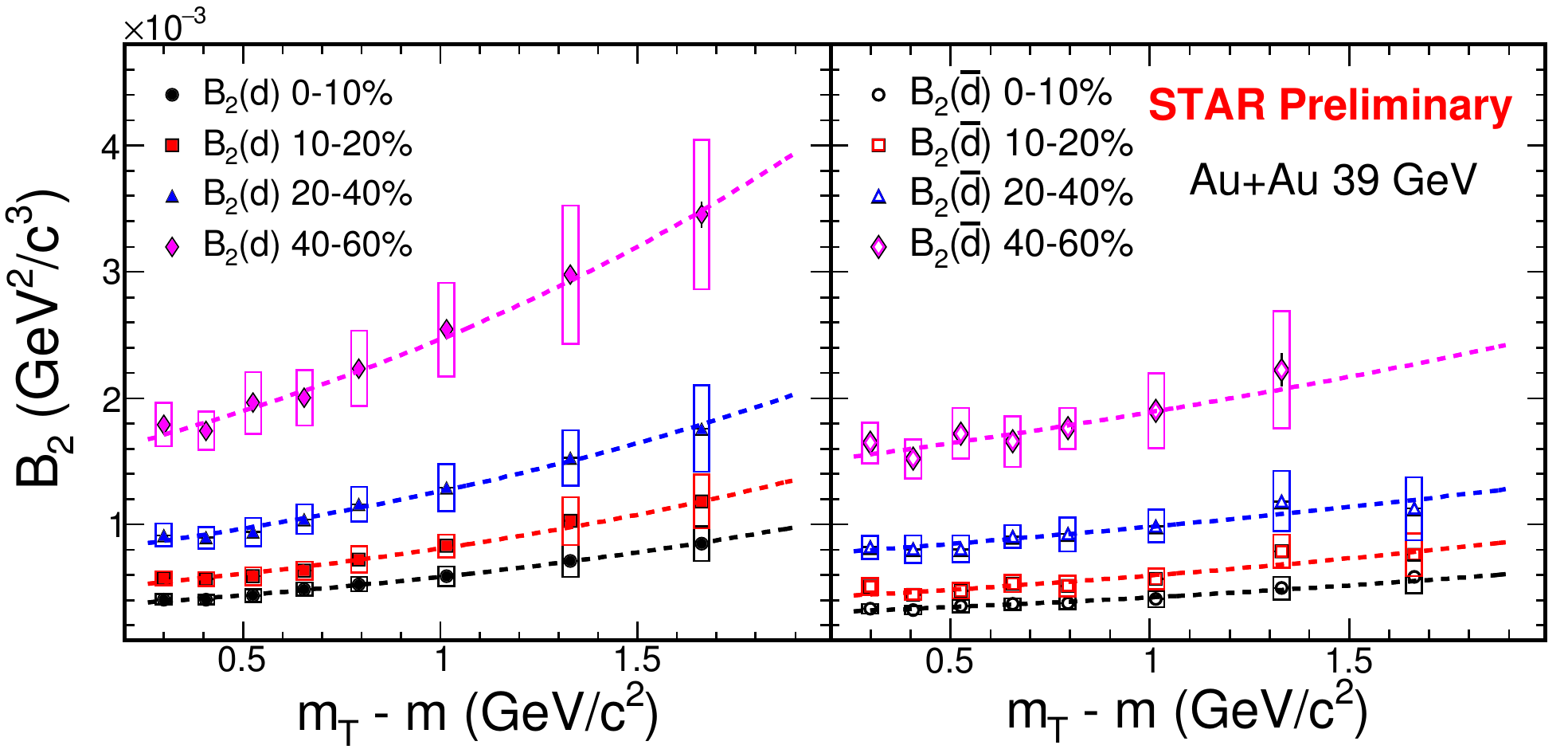}}
\caption{\label{b2pt} Coalescence parameter $B_2$ as a function of $m_t-m$ for deuterons (left panel) and anti-deuteron (right panel) from 39 GeV 0-10\%, 10-20\%, 20-40\%, and 40-60\% Au+Au collisions. The boxes show systematic error and vertical lines show the statistical error separately. The dashed lines represent fits with an exponential in $m_T$.}
\end{figure}

Results of coalescence analysis of deuteron and the proton spectra are presented in Fig.~\ref{b2pt}. This figure demonstrates that the $B_2$ in general depends on $m_T$, which can be described by
\begin{equation}
B_2=a\exp\left[b(m_T-m)\right]
\end{equation}
The $B_2$ increases with $m_T$, which might suggest an expanding collision system. $B_2$ decreases with collision centrality which can be explained by a decreasing source volume, the smaller distance between the protons and neutrons, the more likely they coalesce.

\begin{figure}[!ht]
\centering\scalebox{0.45}[0.45]{\includegraphics{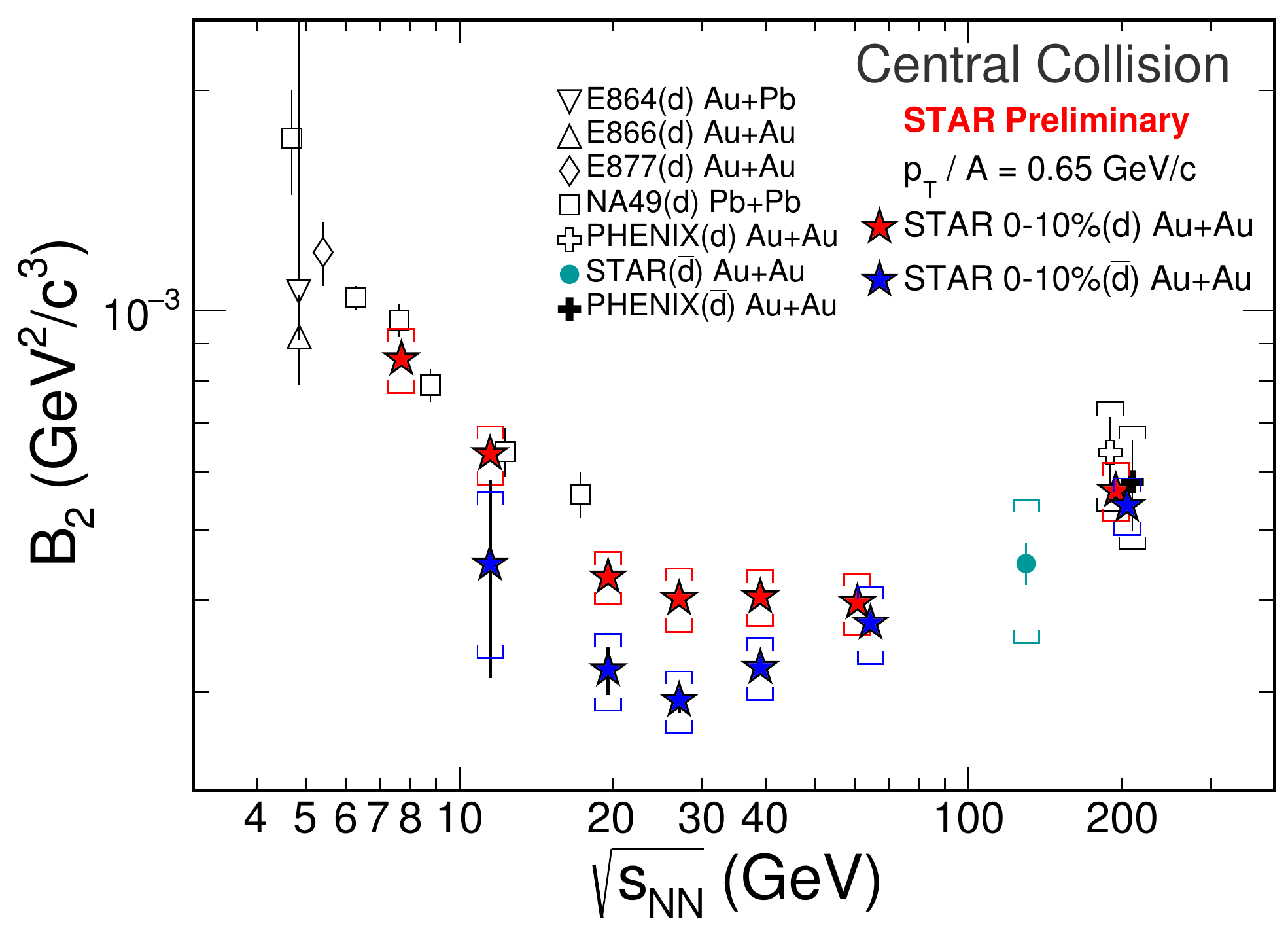}}
\caption{\label{b2} Comparison of the coalescence parameter $B_2$ for deuterons and anti-deuterons
 ($p_T/A =$0.65 GeV/$c$) with other experiments for central A+A collisions. The boxes show systematic error and vertical lines show the statistical error separately.}
\end{figure}

Figure ~\ref{b2} compares the results for 0-10\% $B_2$ ($p_T/A=$0.65 GeV/c) and other experimental central collision data. $B_2$ reaches a minimum around $\sqrt{s_\mathrm{NN}} =$ 20 GeV. Below 20 GeV, the coalescence parameter $B_2$ decreases with $\sqrt{s_{\mathrm{NN}}}$ implied that the emitting source increase with collision energy. These non-monotonic patterns are consistent with the minimum observed for the energy dependence of the viscous coefficients and $\pi$ HBT results~\cite{Adare:2014qvs}. The new observation is that the $B_2(\bar{d})$s are systematically lower than those of $B_2(d)$s implying the emitted source of anti-particles is larger than those of particles.

%\begin{figure}[!ht]
%\begin{minipage}[b]{0.47\linewidth}
%\centering
%\makebox[\linewidth]{%
%\includegraphics[width=0.95\textwidth]{fig/b2pt.eps}}
% \caption{\label{b2pt} Coalescence parameter $B_2$ as a function of $m_t-m$ for deuterons (left panel) and %anti-deuteron (right panel) from 39 GeV 0-10\%, 10-20\%, 20-40\%, and 40-60\% Au+Au collisions. The boxes %show systematic error and vertical lines show the statistical error separately. The dashed lines represent %fits with an exponential in $m_T$.}
%\end{minipage}
%\hspace{0.5cm}
%\begin{minipage}[b]{0.47\linewidth}
%\centering
%\makebox[\linewidth]{%
%\includegraphics[width=1.05\textwidth]{fig/b2.eps}}
%  \caption{\label{b2} Comparison of the coalescence parameter $B_2$ for deuterons and anti-deuterons
% ($p_T/A =$0.65 GeV/$c$) with other experiments for central A+A collisions. The boxes show systematic error %and vertical lines show the statistical error separately.}
%\end{minipage}
%\end{figure}

\section{Conclusions}
We presented systematic studies of $d$ and $\bar{d}$ production in heavy-ion collisions at $\sqrt{s_\mathrm{NN}} =$ 7.7$\sim$200 GeV. The $d/p$ and $\bar{d}/\bar{p}$ ratios can be well reproduced by the thermal model. The isospin extracted from $d/p^2$ is larger than those from $\pi^+/\pi^-$ which may suggest that some of the observed deuteron are from the nuclear fragmentations. The extracted coalescence parameter $B_2$ exhibits a decrease with collision centrality and an increase of transverse mass. $B_2$ from central collision reaches a minimum around $\sqrt{s_\mathrm{NN}} =$ 20 GeV, which might imply a change of equation of state. $B_2(\bar{d})$ values are systematically lower than that of  $B_2(d)$ implying emitted source of anti-baryons is larger than that of baryons.

%% The Appendices part is started with the command \appendix;
%% appendix sections are then done as normal sections
%% \appendix

%%
%% \label{}

%% References
%%
%% Following citation commands can be used in the body text:
%% Usage of \cite is as follows:
%%   \cite{key}         ==>>  [#]
%%   \cite[chap. 2]{key} ==>> [#, chap. 2]
%%

%% References with BibTeX database:

\bibliographystyle{apsrev4-1}
\bibliography{bib}

%% Authors are advised to use a BibTeX database file for their reference list.
%% The provided style file elsarticle-num.bst formats references in the required Procedia style

%% For references without a BibTeX database:

% \begin{thebibliography}{00}

%% \bibitem must have the following form:
%%   \bibitem{key}...
%%

% \bibitem{}

% \end{thebibliography}

\end{document}